\documentclass{article}

\usepackage{graphicx}
\usepackage{amssymb}
\usepackage{amsmath}
\usepackage{stmaryrd}
\usepackage{xspace}
\usepackage{url}
\usepackage{boxedminipage}
\usepackage{subfigure}

\usepackage{bibtopic} 

\newcommand{\codeline}[1]{\texttt{#1}}

\author{Julien Cohen\footnote{LINA (UMR 6241 : CNRS  – Universit\'e de Nantes – \'Ecole des Mines de Nantes)}
 ~~~  
Jean-Louis Giavitto\footnote{IBISC (FRE 3190 : CNRS, Universit\'e d'\'Evry Val d'Essonne, Genopole)} 
~~~ 
Olivier Michel\footnote{LACL (Universit\'e Paris-Est Cr\'eteil)}}

\title{Variable elimination for building interpreters}

\date{}

\begin{document}

\maketitle

\section*{Abstract}

In this paper, we build an interpreter by reusing host
language functions instead of recoding mechanisms of
function application that are already available in the host
language (the language which is used to build the
interpreter). In order to transform user-defined functions
into host language functions we use combinatory logic :
$\lambda$-abstractions are transformed into a composition of
combinators. We provide a mechanically checked proof that
this step is correct for the call-by-value strategy with
imperative features.

\section{Introduction}

When one writes an interpreter for a programming language with
functions, he has to implement a mechanism for reducing function applications.
Moreover, he may need to implement several reduction
mechanisms when there are different kinds of functions in the
language.

Most of the time, the language that is used to build the
interpreter, which we call the \emph{host language}, already
supports functions and their application.
Then is it really necessary to encode in the interpreter such mechanisms
which are already available in the host language? Can't we reuse host language function application instead of recoding it?
Higher order abstract syntax~\cite{HOAS} allows us to do it.
In~\cite{boxes-go-bananas}, an example of such an evaluator
is given, but, as we will see, using it into an interpreter
is generally not direct.

In order not to address a too specific problem, we put the
following constraints to the answer we will give:
\begin{itemize}

\item The technical solution should accept 
  mainstream languages as host language.

\item The interpreted language should possibly have
  higher-order functions, non-strict features and imperative
  features.

\item It should be easy to include in the interpreted language library
  functions, that is functions which are defined in host
  language libraries or in interfaced language libraries.

\item The performances of the interpreter should be
  comparable to other widely used interpreters.

\end{itemize}

In this paper we propose an evaluation scheme which reuses
function application mechanisms of the host language instead
of recoding them.
That scheme is described in section~\ref{sec-scheme-global}
for a functional language with a call-by-value strategy. An
interpreter is built with the OCaml host language.
We will see that the tricky task is to transform
user-defined functions into host language functions. In
order to do this we use a decomposition of user-defined
functions into combinators.
In section~\ref{sec-imperative} we discuss the introduction
of imperative features into the base language considered
previously.
In section~\ref{sec-performances} we show that the performances of the scheme are at least acceptable.
In section~\ref{sec-correctness} we state that the
decomposition of user-defined functions into combinators is
correct for the call-by-value strategy and imperative
features. A mechanically checked proof is provided.
In section~\ref{sec-host-language}, we give clues to implement the scheme with various host languages, and we give examples in Java.
Finally, we discuss related work in section~\ref{sec-conclusion} and we recall the advantages and drawbacks of our approach.

\section{Interpretation with Host Language Reduction}
\label{sec-scheme-global}

In this section, we present our scheme of interpretation by
building an interpreter for a small language.
The host language is OCaml~\cite{caml_site}.
As discussed in section~\ref{sec-host-language}, one can
choose another host language.

We consider as the language to be interpreted a
$\lambda$-calculus with constants, external functions and a
conditional \emph{if-then-else} construct.
This language is described by the following data type \texttt{expr}.

\begin{verbatim}
type expr = 
  | E_Const of value
  | E_Var of ident
  | E_Abs of ident * expr
  | E_App of expr * expr
  | E_If of expr * expr * expr
\end{verbatim}

Constants are handled with the \texttt{E\_Const}
construct. This constructor is used to represent integer constants, boolean constants and external functions.
We use the type \texttt{value} defined below to represent constants.

\begin{verbatim}
type value = 
  | V_Int of int
  | V_Bool of bool
  | V_Fun of (value->value)
    ...   
\end{verbatim}

External functions are represented by \texttt{value->value} host language functions.
They are embedded in the type \texttt{value} with the \texttt{V\_Fun} construct.

In order to compare our idea to the classical setting,
we first build a classical interpreter  using an
environment in order to deal with $\lambda$-abstraction parameters (section~\ref{sec-classical}).
After that, we will present our interpretation scheme (section~\ref{sec-scheme}).
The type \texttt{value} is a bit different in these two
schemes. These is why we let some dots in its
definition given above.
We will then discuss the performances of our scheme (section~\ref{sec-performances-1}).

\subsection{A classical scheme}
\label{sec-classical}

We build here a simple call-by-value interpreter with the
classical use of environments. We hide the implementation of
environments, a naive approach would be to use
association lists.

\begin{verbatim}
type 'a environment
val env_assoc : ident -> 'a environment -> 'a
val env_add   : ident -> 'a -> 'a environment
\end{verbatim}

We have to add one more construct to the type of the values given sooner:
user defined functions (\emph{i.e.} $\lambda$-abstractions
by opposition to external/pre-defined functions), which are represented by
closures. Closures are $\lambda$-abstractions equipped with the
environment in which they have been defined.

\begin{verbatim}
type value = 
  | V_Int of int
  | V_Bool of bool
  | V_Fun of (value->value)
  | V_Closure of ident * expr * (value environment)   
\end{verbatim}

This abstract syntax does not contain variables, applications,
or conditionals. 
However, such constructs may appear in the body of a
closure.
The types \codeline{value} and \codeline{expr} are mutually recursive (they should be declared together in OCaml).

The type \codeline{value} contains two kinds of functions :
$\lambda$-abstractions represented by closures and external
functions.
Each kind of function will have a particular reduction mechanism.
Let us have a look at these mechanisms in the evaluation function. 
That function takes as parameters an environment and the
expression to evaluate.

\begin{verbatim}
let rec eval env = function
  | E_Const v -> v
  | E_Var i -> env_assoc i env
  | E_Abs (i,e) -> V_Closure (i,e,env)

  | E_App (e1,e2) ->
      ( match (eval env e1, eval env e2) with
          | V_Closure (x,e,env2), v2 -> eval (env_add x v2 env2) e
          | V_Fun f, v2 -> f v2      (* caml application here *)
          | _ -> raise TypeError                                  ) 

  | E_If (e1,e2,e3) ->
      ( match (eval env e1) with
          | V_Bool true  -> eval env e2
          | V_Bool false -> eval env e3
          | _ -> raise TypeError                                  )
\end{verbatim}

The reduction of external function applications uses a
simple host-language application of the function to
its parameter.
The reduction of $\lambda$-abstraction applications builds a
new environment and evaluates the body of the abstraction,
achieving an usual $\beta$-reduction.

We can see that the applications of
host language functions are more easily expressed than
applications of $\lambda$-abstractions since their reduction is delegated to the host
language.

\subsection{Our interpretation scheme}
\label{sec-scheme}

We are now going to present our proposal to build a simpler
interpreter.
In order to do this, we are going to unify the different
kinds of functions that are handled by transforming
$\lambda$-abstractions into host-language functions.
First, let us see why this unification yields a simpler
evaluator.

\subsubsection{The core evaluator}

We would like to use the
 language \texttt{m\_expr} (\emph{minimal expressions}) to represent expressions:

\begin{verbatim}
type m_expr = 
  | M_Const of value
  | M_App of  m_expr *  m_expr
\end{verbatim}

This language contains host language functions in
the \codeline{M\_Const} construct, but
it does not contains abstractions anymore.
Furthermore, it does not contains variables since variables
cannot be bound by abstractions anymore.
We have also removed the conditional construct \codeline{E\_If} from the
language \codeline{m\_expr}, assuming it can be transformed
into function applications.
This is discussed in section~\ref{sec-non-strict}

Additionally, we do not need
closures in the type \codeline{value} anymore (and \codeline{value} does not depend on the type \codeline{expr} anymore):

\begin{verbatim}
type value = 
  | V_Int of int
  | V_Bool of bool
  | V_Fun of (value -> value) 
\end{verbatim}

At this point, as we expect, 
the \texttt{eval} function is extremely simple : there is only
one kind of application which reduction is performed by the
host language.
Moreover, there is no need for dealing with environments.

\begin{verbatim}
let rec eval = function 
  | M_Const v -> v
  | M_App (e1, e2) -> 
      ( match (eval e1, eval e2) with
           | (V_Fun f , v) -> f v (* caml application *)
           | _ -> raise TypeError                      ) 
\end{verbatim}

\subsubsection{Our scheme of interpretation}

Two steps are necessary to reach the type \texttt{m\_expr} from \texttt{expr} : \begin{enumerate}

\item
transforming $\lambda$-abstractions into host language
functions, which we call \emph{variable elimination}, and

\item
transforming conditionals into functions, which we call
\emph{non-strictness elimination}\footnote{Non-strictness
elimination could also be performed in the classical setting
in order to simplify the \texttt{eval} function, but it is
more natural in our scheme where we tend to transform most
constructs into functions.
}.
\end{enumerate}
Our interpretation scheme is pictured in the following diagram where
$\mathcal P$ stands for the non-strictness elimination,
$\mathcal C$ stands for the variable elimination and
$\mathcal E$ stands for the evaluation step.

$$\mathtt{expr}    \xrightarrow{\mathcal P} 
  \mathtt{p\_expr} \xrightarrow{\mathcal C}
  \mathtt{m\_expr} \xrightarrow{\mathcal E} \mathtt{value}$$

We use a different type for the result of each step in order to show
that each step is independent from the others.

We now explain the first two steps, $\mathcal P$ and~$\mathcal C$, in sections~\ref{sec-non-strict} and~\ref{sec-ski}.

\subsubsection{Non-strictness elimination}
\label{sec-non-strict}

The \emph{if-then-else} conditional construct is a typical
example of non-strict function, that is a function that does
not always evaluate its arguments (the three branches of the
conditional are seen as its arguments).
Pattern matching of ML, \emph{switch/case} construct of C,
 macros of LISP, and lazy logical operators are other
 examples of non-strict functions. 
Even the sequence of imperative languages can be seen as a
 non-strict function.

We want to transform
non-strict functions into standard call-by-value functions.
The classical way to suspend the evaluation of the
argument of a function in a call-by-value setting is to
encapsulate it into an abstraction whose body will be
evaluated when necessary~\cite{thunks}.
This is the solution we choose here.
The terms after non-strictness elimination are in the
\texttt{p\_expr} type (\emph{purely functional expressions}) :
\begin{verbatim}
type p_expr = 
  | P_Const of value
  | P_Var of ident
  | P_Abs of ident *  p_expr
  | P_App of p_expr *  p_expr
\end{verbatim}

The non-strictness elimination, $\mathcal P$ in the diagram, is given by
the \texttt{purify} function which transforms only the
\emph{if-then-else} construct.
\begin{verbatim}
let rec purify = function 
  | E_Const c -> P_Const c
  | E_Var x -> P_Var x
  | E_Abs (x,e) -> P_Abs (x, purify e)
  | E_App (e1, e2) -> P_App (purify e1, purify e2)
  | E_If (e1, e2, e3) ->  
      let e'1 = purify e1
      and e'2 = P_Abs(dummy_var, purify e2)
      and e'3 = P_Abs(dummy_var, purify e3)
      in P_App(P_App(P_App (P_Const (V_Fun fun_IF),e'1),e'2),e'3) 
\end{verbatim}

The behavior of the conditional is realized at evaluation time by the function
\texttt{fun\_IF} given below.
That function checks its first argument (a value)
and triggers the evaluation of the relevant \emph{then}
or \emph{else} branch by applying it to a dummy parameter.
That application is reduced by the host language.
 The expressions \texttt{e'2} and \texttt{e'3} above being
 abstractions, they will be transformed into host
 language functions (or function expressions, that is expressions evaluating into functions) by the variable elimination step.
 This
 explains why the function \texttt{comb\_IF}  expects
 \texttt{e2} and \texttt{e3} to be functions.

\begin{verbatim}
let fun_IF =  
  fun e1 ->  V_Fun (
    fun (V_Fun e2) -> V_Fun (
      fun (V_Fun e3) ->
        match e1 with
          | V_Bool true  -> e2 dummy_val (* caml application *)
          | V_Bool false -> e3 dummy_val (* caml application *)
          | _ -> raise TypeError                             ))
\end{verbatim}

\subsubsection{Variable elimination}
\label{sec-ski}

\newcommand{\combii}{\mathsf I}
\newcommand{\combis}{\mathsf S}
\newcommand{\combik}{\mathsf K}
\newcommand{\combib}{\mathsf B}
\newcommand{\combic}{\mathsf C}
\newcommand{\combin}{\mathsf N}

The goal of this step is to transform terms with variables
and $\lambda$-abstractions (type \texttt{p\_expr}) into terms
of the \texttt{m\_expr} type, that is, without variables and
abstractions.

In multi-level programming languages, that step is done
directly by creating at run-time new expressions of the host
language that can be used immediately\footnote{\label{ex-scheme}For
instance, the Scheme expression \texttt{(eval (list 'lambda (list
(string->symbol "x")) (list '+ (string->symbol "x") 1) ))}
allows to create the Scheme function \texttt{(lambda (x) (+ x 1))}.
}.
But few languages provide this kind of features and we will
do without it.

Combinatory logic~\cite{combinatory-logic,
history_of_lambda_calculus} provides many algorithms to
translate $\lambda$-terms into compositions of pre-defined
functions called combinators. We will naturally find
candidates for our variable elimination step among these
algorithms.
However, one should remember 
that most of these algorithms are designed for the
call-by-name evaluation strategy whereas we are interested in call-by-value.

We illustrate our scheme by using a translation into 
compositions of combinators S, K and I.
The classical algorithm applies the rewriting rules (a), (b) and (f) of figure~\ref{fig-elim-ski}
in the following order : (f) wherever possible, then
(a) then (b).
It is noted (fab).

\begin{figure}[ht]
\begin{align*}
\lambda x . U & \rightarrow (\combik ~ U) & (a)\\
\lambda x . x & \rightarrow \combii       & (b)\\
\lambda x.(M~N) & \rightarrow ((\combis ~ \lambda x. M) ~ \lambda x.N) & (f)\\[2mm]
\lambda x . c & \rightarrow (\combik ~ c) & (a_v)
\end{align*}
\begin{align*}
\combii & = \lambda x.x\\
\combik & = \lambda x. \lambda y.x\\
\combis & = \lambda x.\lambda y. \lambda z . ( (x\,z)\,(y\,z))
\end{align*}

\caption{Basic variable elimination rules and combinators}
\label{fig-elim-ski}
\end{figure}

In theses rules, $M$ and $N$ stand for any term while $U$
stands for a term where $x$ does not appear (more precisely,
$x$ does not appear in a free position). 
Since (a) is applied after (f), $U$ may not contain
applications.
 Therefore, in the (fab) algorithm one can replace the (a)
 rule by the rule (a$_v$) of figure~\ref{fig-elim-ski} where $c$ is a
 constant, an external function or a variable different from
 $x$.

In fact, the rule (a) taken alone is not correct in the
call-by-value setting whereas ($a_v$) is.
For this reason we use the rules (f), ($a_v$)
and (b), in any order (they are disjoint).

That transformation is realized by the
function \texttt{elim} given below.

\begin{verbatim}
let rec elim = function
    P_Const c    -> P_Const c
  | P_Var   x    -> P_Var x

  | P_App (e1, e2) ->  P_App (elim e1, elim e2) 
      
  | P_Abs(s, P_Var s' ) -> 
      if s = s' 
      then  cI 
      else  (P_App (cK, P_Var s'))

  | P_Abs(_, P_Const c) -> P_App (cK, P_Const c )

  | P_Abs(s,P_App(e1,e2)) ->
       P_App(P_App(cS, elim (P_Abs(s,e1))), elim (P_Abs(s,e2)))
      
  | P_Abs(s, (P_Abs(_,_) as t))->  elim (P_Abs(s,elim t))
\end{verbatim}

As the \texttt{elim} function has to apply to its result, its
return type has to be \texttt{p\_term} as the type of its
argument.
For this reason, we use the following
\texttt{check\_no\_var} function to transform
\texttt{p\_expr} terms into \texttt{m\_expr} terms after
\texttt{ski} has been applied~:

\begin{verbatim}
let rec check_no_var = function
  | P_Const c -> M_Const c
  | P_App (e1, e2) -> M_App (check_no_var e1, check_no_var e2)
  | P_Var _ -> raise UnexpectedError
  | P_Abs _ -> raise UnexpectedError

let trans e = check_no_var (elim e) 
\end{verbatim}

If the term passed to \texttt{elim} does not contain any free
variables then the transformed term does not contain any
variable nor abstraction and the \texttt{check\_no\_var}
function cannot fail.

The combinators $\combis$, $\combik$, and $\combii$ used in \texttt{elim} are defined as follows :
\begin{verbatim}
let cI = P_Const (V_Fun (fun x -> x)) 

and cK = P_Const (V_Fun (fun c -> V_Fun (fun _ -> c)) )

and cS = P_Const (V_Fun ( 
  fun (V_Fun f) -> V_Fun (
    fun (V_Fun g) -> V_Fun ( 
      fun x -> match (f x) with       (* 1 caml capplication *) 
          V_Fun h -> h (g x)     )))) (* 2 caml applications *) 
\end{verbatim}
\label{code-definition-combinators}

\subsubsection{Handling of higher-order library functions}

Let us focus on the handling of higher-order functions in
our scheme. Let us suppose we want to provide access to an
higher-order function defined in the host language. We take
as an example the function \codeline{host\_compose} defined
as follows:

\begin{verbatim}
let host_compose f g x = f (g x)
\end{verbatim}

In order to give access to this function in the interpreted
language, we have to provide a wrapper to transform it into
a \codeline{value->value} function and embed it into
a \codeline{V\_Fun} constructor. 
Here is such a wrapper:

\begin{verbatim}
let compose_embed = 
  V_Fun (function (V_Fun f) -> 
     V_Fun (function (V_Fun g) ->
        V_Fun (host_compose f g) )) 
\end{verbatim}
Since functions in the interpreted language are represented
by host language functions, the wrapper is very simple: it
just unwraps the two arguments received and pass them
to \codeline{host\_compose}.

In the classical approach described sooner, the wrapper
would need to make the difference between external functions
and $\lambda$-abstractions. The wrapper
for \codeline{host\_compose} could be as follows :

\begin{verbatim}
let app f v = match f with
  | V_Fun h -> h v 
  | V_Closure (x,e,env) -> eval (env_add x v env) e

let compose_embed = 
  V_ExtFun (function f -> 
     V_ExtFun (function g -> 
        V_ExtFun (host_compose (app f) (app g) )))
\end{verbatim}
Here we have a dynamical dispatch to host language
application (case \codeline{V\_Fun}) or to interpreted
application reduction (case \codeline{V\_Closure}).


We have now detailed all the steps of our interpretation
scheme. 
However, the \texttt{elim} variable elimination algorithm (fab/fa$_v$b) is
not plainly satisfactory because, as is well known, it creates terms very
inefficient to evaluate.
We investigate other variable elimination rules in section~\ref{sec-performances-1}.

\subsection{Towards better performances}
\label{sec-performances-1}

The use of the combinators $\combis$, $\combik$ and $\combii$
leads to an explosion of the size of the terms, and thus to
an explosion of the evaluation time.
We propose here additional rewriting rules to improve the size (and the
evaluation speed) of the terms produced.
All the rules we present fit 
exactly in our scheme, that is they can be added to the
variable elimination algorithm without need to modify the
other steps.
This way we keep the simplicity of the whole scheme and the
simplicity of the evaluation function.

\subsubsection{Using the $\combib$, $\combic$ and $\combin$ combinators}
\label{sec-BCN}

\begin{figure}[ht]
\begin{align*}
\lambda x . (U~M) & \rightarrow ((\combib ~ U) ~ \lambda x.M)  & (d)\\
\lambda x . (M~U) & \rightarrow ((\combic ~ \lambda x .M)~ U)  & (e)\\
\\
\lambda x . (c~M) & \rightarrow ((\combib ~ c) ~ \lambda x.M)  & (d_v)\\
\lambda x . (M~c) & \rightarrow ((\combic ~ \lambda x .M)~ c)  & (e_v)\\
\lambda x . (c_1~c_2) & \rightarrow ((\mathsf N ~ c_1)~ c_2)  & (n_v)\\
\end{align*}
\begin{align*}
\combib & = \lambda a.\lambda g. \lambda x. (a~(g~x))\\
\combic & = \lambda f.\lambda b. \lambda x. ((f~x)~b)\\
\combin & = \lambda a.\lambda b. \lambda x. (a~b)
\end{align*}
\caption{Selective rules and combinators}
\label{fig-rules-bcn}
\end{figure}

The combinators $\combib$ and $\combic$ are classically used with the rules
 (d) and (e) of figure~\ref{fig-rules-bcn} where $U$ does
 not contain $x$ free but $M$ may.
These two rule are not correct in call-by-value when $U$
contains applications because it leads to evaluation of
applications that are initially protected in the body of a $\lambda$-abstraction.
For this reason, we use the rules (d$_v$) and ($e_v$) instead of~(d) and~(e).
 In addition, we propose to use the rule (n$_v$) together with
 (d$_v$) and ($e_v$). 
That rule follows the same idea as (d$_v$) and ($e_v$) to
 avoid the distribution of a $\lambda$ over branches of an
 application.
We use the standard combinators $\combib$ and $\combic$ and a third
combinator noted~$\combin$ (Fig~\ref{fig-rules-bcn}).

\subsubsection{Safely reducing partial applications before evaluation time}
\label{sec-pre-eval}

The form $\lambda x.c$ is transformed into $(K\,c)$ by the rule (a) or (a$_v$). 
Reducing the application of K to $c$ before evaluation time
does not change the semantics of the term since we do not
reduce an application that was in the initial term but one
which has been artificially added.
At the risk of spending time reducing an application that might be never
reduced at evaluation time, we can reduce once an
application that may be reduced many times at evaluation
time, leading to a speedup.

This kind of pre-evaluation can be
considered whenever a combinator is introduced.
In fact, given a combinator of the form $\lambda x_1.\lambda
x_2. \dots \lambda x_n . E$, we can safely reduce its
applications to $n-1$ arguments as long as these
arguments are values or can be safely pre-evaluated into
values.

All the combinators we have presented have this
form. Moreover, the rules that introduce them apply them to
``$n-1$'' arguments.

We present two ways to apply this idea : ``by hand'' by
providing rules introducing pre-reduced forms or
automatically by launching a kind of eval when we know it is
safe.

\paragraph{Hand-made reduction at transform time.}

Let us consider again the example of $\lambda x.c$
transformed into $(\mathsf K ~c)$ by the rule $(a_v)$.  The following line in the function \codeline{elim} given
above does this transformation (except when $c$ is a variable):
\begin{verbatim}
  | P_Abs(_, P_Const c) -> P_App (cK, P_Const c )
\end{verbatim}

We can compute by hand the result of evaluation of the expression \codeline{P\_App (cK, P\_Const c)}. 
It is \codeline{V\_Fun (fun \_ -> c)} (see the definition of \codeline{cK} on page~\pageref{code-definition-combinators}).
So we can produce directly the
convenient host language function with :

\begin{verbatim}
  | P_Abs(_, P_Const c) -> P_Const (V_Fun (fun _ -> c)) 
\end{verbatim}

\newcommand{\combikc}{\mathsf K _c}

This is formalized  with the rule ($\tilde{a_v}$) where $c$ is a constant or an external function (not a variable because the created function cannot be explored by further transformation steps) and $\combikc$ is defined by $\lambda x.c$ :

\begin{align*}
\lambda x . c & \rightarrow \combikc & (\tilde a_v)
\end{align*}

It can seem strange that $\lambda x.c$ is transformed into
$\lambda x.c$ (the $x$ is not necessarily the same in the
two terms) but the second term is implemented by a host
language function whereas the first one is not.

Here, $\combikc$ is not a combinator but a family of combinators indexed by $c$. We can generate each of them with the following function :
\begin{verbatim}
let make_K_c c = P_Const (V_Fun (fun _ -> c)) 
\end{verbatim}
Each necessary combinator will be created at transform time,
possibly twice.

We can apply this kind of by-hand pre-evaluation wherever the rules $(a_v)$, $(d_v)$, $(e_v)$ and $(n_v)$ apply. Here are the corresponding rules: 

\begin{align*}
\lambda x . (c~M) & \rightarrow (\mathsf B_c  ~ \lambda x.M)  & (\tilde {d_v})\\
\lambda x . (M~c) & \rightarrow (\mathsf C_c ~ \lambda x .M)  & (\tilde {e_v})\\
\lambda x . (c_1~c_2) & \rightarrow \mathsf N_{c_1 c_2}  & (\tilde {n_v})\\
\end{align*}

The rule $(f)$ introducing the combinator $\combis$ is not
concerned since if one of the term is a constant, the rules
$(d/\tilde {d_v})$ or $(e/\tilde {e_v})$ will be preferred.

\paragraph{About dynamic generation of combinators.} 
One may wonder why we cannot generate at transform time the
host-language function corresponding directly to an
expression since we are able to generate ad-hoc combinators.

The reason is that we cannot make the link between the
parameter of a function and the corresponding bound variable
at an arbitrary depth in an expression.
Indeed, to be able to handle expressions of arbitrary depth,
we would need for instance a recursive operation and the
host-language binding could not be preserved during the
recursive calls (except by achieving a kind of environment
but that environment would need to be accessed at execution
time, which we want to avoid in this work).

For this reason, we cannot generate at transform time a
host-language function for each expression.  So we are
restricted to do it for most frequent forms and for the
general form we use the variable elimination rules given
above.

\paragraph{Automatic reduction at transform time.}
\label{sec-auto-pre-reduce}
We have seen how we can modify the transformation rules to
achieve some pre-evaluation. A similar result can be obtained by
systematically reducing partial applications of combinators  before
evaluation time.
To avoid to change the semantics of the terms, we have to
check which applications can be pre-reduced.

Let us consider the combinator $\combis$.
It waits for 3 arguments and none application reduction is
triggered before the 3 arguments are received.
Let us now consider the form $(((\combis ~ v_1) ~ v_2)$, where $v_1$
and $v_2$ are constants or external functions.
For the reason given above, we can safely reduce the two
applications without risking to apply $v_1$ or $v_2$.

For any combinator which does not apply anything before
having received all its arguments, each time we produce a
form where it misses an argument we can pre-evaluate that
form. We do this with the following function which checks that the
arguments are values or can safely be pre-reduced into values:

\begin{verbatim}
let rec pre_eval_app e = match e with
  | P_App (e1,e2) ->
      ( match (pre_eval_app e1, pre_eval_app e2) 
        with 
          | (P_Const (V_Fun f),P_Const v) -> P_Const (f v)
          | (e'1,e'2) -> P_App (e'1,e'2)
      )
  | _ -> e
\end{verbatim}
 
This allows to have a general mechanism for pre-evaluation
instead of using ad-hoc rules where pre-evaluations have
been made by hand.
In our transformation function \codeline{elim}, we can
apply \codeline{pre\_eval\_app} each time we generate an
expression. For instance, here is the difference in \codeline{elim} for the
introduction of the combinator $\combis$:

\emph{Code without pre-evaluation:}
\begin{verbatim}
  | P_Abs(s,P_App(e1,e2)) ->
       P_App(P_App(cS, elim (P_Abs(s,e1))), elim (P_Abs(s,e2)))
\end{verbatim}

\emph{Code with automatic pre-evaluation:}
\begin{verbatim}
  | P_Abs(s,P_App(e1,e2)) ->
       pre_eval_app(
        P_App(P_App(cS, elim (P_Abs(s,e1))), elim (P_Abs(s,e2)))
        )
\end{verbatim}

The same can be done for all the rules we have proposed so far. 

Doing pre-evaluation during transformation time gives better results than doing it between transformation and evaluation because applying \codeline{elim} on simpler terms may allow to use a different combinator (for instance a $\combik$ instead of an~$\combis$).

\subsubsection{Using a combinator dedicated to the conditional construct} 
\label{sec-sif}

\newcommand{\combisif}{\mathsf S _\textit{IF}}

When the conditional is under an
abstraction, that abstraction is going to be distributed over
the three branches of the conditional with the variable elimination 
algorithms presented so far. 
Thus it may not be necessary to over-protect the \emph{then}
and \emph{else} branches with additional dummy abstractions in this case.
This is the idea we develop in this section. 
In order to be able to dump the dummy abstraction, we use a
dedicated combinator noted $\combisif$ which will selectively
distribute the received value onto the convenient branch.
The corresponding rewriting rule is the following :

\begin{align*}
\lambda x.(((\mathit{IF} ~ e_1) ~ \lambda z.e_2) ~ \lambda z.e_3) & \rightarrow
(((S_\textit{IF} ~ \lambda x.e_1) ~ \lambda x.e_2) ~ \lambda x. e_3)
 & (i)
\end{align*}

In this rule, the dummy $\lambda z$ abstractions introduced
during the non-strictness elimination step are discarded.
The combinator $\combisif$ plays the role of the $\combis$ and
$\mathit{IF}$ (\texttt{fun\_IF} in section~\ref{sec-non-strict}) combinators.
It is defined as follows :

\begin{verbatim}
let cSIF =  P_Const (
    V_Fun (fun (V_Fun e1) -> 
      V_Fun (fun (V_Fun e2) ->
        V_Fun (fun (V_Fun e3) ->
           V_Fun (fun x ->
              match e1 x with         (* caml application *)
                V_Bool true  -> e2 x  (* caml application *)
              | V_Bool false -> e3 x  (* caml application *)
              | _ -> raise TypeError                        )))))
\end{verbatim}

\subsubsection{Multiple abstractions and multiple applications}
\label{sec-n-ary}

Patterns like $\lambda x. \lambda y. (e_1 ~ e_2)$ and
$\lambda x. (e_1 ~ e_2 ~ e_3)$ often occur in real
programs or in temporary expressions encountered during the
variable elimination rewriting steps.
In this section, we introduce rules for dealing with several abstractions
over an application, with an abstraction over several
applications and other combinations of these kinds.

Let us consider the form  $\lambda x. \lambda y. (e_1 ~ e_2)$.
The $(fa_vb)$ algorithm would transform it into  $(\combis~((\combis~(\combik~\combis))~[\lambda x.\lambda y.e_1])~[\lambda x . \lambda y.e_2])$ (the brackets denote the application of the variable elimination algorithm considered).
We propose instead the rule $(s^2)$ given in figure~\ref{fig-rules-2}. 
The corresponding combinator $\combis^2$ is also given in figure~\ref{fig-rules-2}. 
Using $(s^2)$ in the considered example produces a form with
2 applications instead of a form with 5 applications for the
$(fa_vb)$ algorithm.

\begin{figure}[ht]
\begin{align*}
\lambda x. \lambda y. (e_1 ~ e_2) &
 \rightarrow ((\combis^2 ~\lambda x.\lambda y.e_1)~\lambda x . \lambda y.e_2) &
 (s^2)\\
\lambda x. (e_1 ~ e_2 ~ e_3) &
\rightarrow (((\combis_2~\lambda x.e_1) ~\lambda x.e_2)~\lambda x. e_3) &
(s_2)\\
\lambda x.\lambda y.(e_1 ~ e_2 ~ e_3) &
\rightarrow (((\combis_2^2 ~\lambda x.\lambda y.e_1)~\lambda x . \lambda y.e_2) ~\lambda x.\lambda y.e_3 &
(s_2^2)\\
\lambda x.\lambda y . c &
\rightarrow (\combik^2~c) &
(k^2)\\
\lambda x.\lambda y.x &
\rightarrow \combii^2 &
(i^2)
\end{align*}

\begin{align*}
\combis^2 & =  \lambda f.\lambda g.\lambda x.\lambda y. ((f~x~y)~(g~x~y))\\
\combis_2 & =  \lambda f.\lambda g.\lambda h.\lambda x.((f~x)~(g~x)~(h~ x))\\
\combis_2^2 & = \lambda f.\lambda g.\lambda h.\lambda x.\lambda y.((f~x~y)~(g~x~y)~(h~ x~y))\\
\combik^2 & = \lambda x.\lambda y.\lambda z.x\\
\combii^2 & = K
\end{align*}

\caption{Rules and combinators for multiple abstractions/applications}
\label{fig-rules-2}
\end{figure}

Now, let us consider the form $\lambda x. (((e_1 ~ e_2) ~ e_3)$.
 Instead of transforming it into $((\combis~((\combis~[\lambda x.e_1]) ~[\lambda x.e_2]))~[\lambda x. e_3])$,
 we propose the rule $(s_2)$ of figure~\ref{fig-rules-2}.
Here we produce 3 applications instead of 4.

We can also combine these two rules into the rule $(s_2^2)$
which is better than applying separately the two rules. The
use of $\combik$ and $\combii$ can also be extended with the
rules ($k^2$) and ($i^2$).
Of course, these rules can be extended for more than two
abstractions and applications. 
However, it should be noted that as the number of
combinators increases, the number of cash misses at the
processor level during evaluation also increases, which can
lead to decreasing performances.
For this reason, some optimizations are interesting only for
some processors and it is difficult to predict the result of
a particular optimization.

Generalizing the use of the combinators $\combib$, $\combic$
and $\combin$ to multiple abstractions and applications
gives the family of combinators described by Diller~\cite{diller02}.
However, as we have restricted the use of $\combib$ and
$\combic$ to fit with a call-by-value strategy, the use of
Diller combinators has to be restricted in the same way.

\subsubsection{On the complexity of the code}
As we add new rewriting rules and combinators, the code of
the interpreter becomes more complex to handle.
This seems to be in opposition with our goal to have a
simple scheme.
It is true that the variable elimination step becomes more complex as we want to optimize the produced code. However :
\begin{itemize}

\item
  Since the three steps steps of the scheme are independent,
  the optimization of this step does not affect the code of
  the other steps. In particular, the \codeline{eval} function (step $\mathcal E$) is left clean.

\item 
When you add features in the interpreted language, you do
not have to modify the variable elimination step because all
the features are transformed into functions by the previous
step ($\mathcal P$).

\item 
The algorithm is defined by independent rules, which should
make easy to have a well structured code. Then it should be
easy to add new rules to the algorithm.
For instance, for the OCaml code given here, adding a rule
is done by adding a case to a structural pattern matching.

\end{itemize}

For these reasons, we believe that the optimization of this step is compatible with our goal of simplicity.

\subsubsection{Conclusion on the need to optimize}
\label{sec-performances-2}

We have proposed in this section several improvements of the variable
elimination algorithm. 
As we will see in
section~\ref{sec-performances}, a few of these improvements
are sufficient to cope with the explosion of the size of the
produced terms and to provide performances comparable to a naive classic approach.

We will also see that the presented optimizations are not
sufficient to provide guaranteed drastic speedups compared
to the naive classic approach.
In order to achieve this, one should consider other families
of optimizations, including well known ones.

In the next section, we see that imperative features fit well in our scheme.

\section{Imperative Features}
\label{sec-imperative}

In this section, we illustrate the handling of imperative
features in our scheme.
In order to do this, we add to our language two operators, \emph{get}
and \emph{set}, to read and write into references
(\emph{i.e.} imperative variables):

\begin{verbatim}
type expr = 
  ...
  | E_Get of tag
  | E_Set of tag * expr
\end{verbatim}

The type \codeline{tag} is used to name references.
A term \texttt{E\_Get(t)} stands for a read access
to the reference named \texttt{t} and \texttt{E\_Set(t,e)} stands for the
writing of the \emph{value} of \texttt{e} into \texttt{t}.
The complete semantics is given by the classical evaluator
implementation in section~\ref{sec-classical-imperative}.
References are not values here : they cannot be handled
independently of a \emph{get} or \emph{set} operator.

The values associated to references at a given moment are
gathered into a store.
The structure of stores is left abstract.

\begin{verbatim}
type 'a store 
value store_get : tag -> 'a store -> 'a
value store_set : tag -> 'a -> 'a store
\end{verbatim}

We first discuss different possibilities to take imperative
features into account, we illustrate two of them with the
classical interpreter, and then we show how to integrate it
into our scheme.

\subsection{Dealing with imperative features}
\label{sec-classical-imperative}

There are several ways to implement our operators.

\paragraph{The monad approach.}
One purely functional approach is to see stores as values of
the language and to pass the store as an argument to
functions which have to access to it (see~\cite{monades93}).
To modify the store, a function has to return a
new store which will be taken as argument by an other
function.

Taking the store as a parameter and returning it creates a
data flow which enforces the control. It is very useful in
lazy languages where the control is difficult to anticipate.

This approach has the advantage that it fits very well in
purely functional calculus : imperative features become
purely functional. 
For this reason, it would fit directly in
our scheme. 
However, we are going to show that a less purely functional
approach fits too and we will not deal anymore with the monad approach in the rest of this section.

\paragraph{Another functional approach.}

Another approach is to let the eval function handle the
store which is no longer a value of the language but rather
a parameter of the evaluation.
Unlike the environment (see section~\ref{sec-classical})
which is independent in the two branches of an application, the
store resulting of the evaluation of one of the branches of an
application must be passed to the evaluation of the second
branch.

The classical evaluator can be modified as follows to apply this idea.


\begin{verbatim}
let rec eval env s = function
  | E_Const v -> (v, s)
  | E_Var x -> (env_assoc x env, s)
  | E_Abs (x,e) -> (V_Closure (x,e,env), s)

  | E_App (e1,e2) ->
     let (v2,s2) = eval env s e2 (* we have to choose an order
                                     of evaluation to pass the
                                     store between the two evaluations *)
      in ( match eval env s2 e1 with
             | (V_Closure (x,e,env2), s') -> eval ( env_add x v2 env2 ) s' e
             | (V_ExtFun f, s') -> (f v2, s')  (* caml application *)
             | (_,_) -> raise TypeError                              
         ) 
	  
  | E_If (e1,e2,e3) ->
      (
        match eval env s e1 with
            (V_Bool true, s')  -> eval env s' e2
          | (V_Bool false, s') -> eval env s' e3
          | _ -> raise TypeError                                 )

  | E_Get t -> let v = store_get t s in (v, s)
  | E_Set (t,e) -> let (v, s') = eval env s e in 
      (v, store_set t v s')  
\end{verbatim}

Note that eval returns a value and a store.

\paragraph{An imperative implementation.}
This extension of the classical evaluator is more complicated
than the original one.
A way to simplify it is to consider the store as a mutable
global variable.
If we do so, we can still have confidence in the use of the
store because \texttt{eval} is the only function to access
it, it is (or it should be) sequential and the accesses are
triggered at definite times.
Here is such an implementation:


\begin{verbatim}
let rec eval env = function
  | E_Const v  -> v
  | E_Var x -> env_assoc x env 
  | E_Abs (x,e) -> V_Closure (x,e,env)

  | E_App (e1,e2) ->
      (match (eval env e1, eval env e2) with 
         | V_Closure (x,e,env2), v2 -> eval ( env_add x v2 env2 ) e
         | V_ExtFun f , v2-> f v2  (* caml application *)
         | _ -> raise TypeError                              
        ) 
                
  | E_If (e1,e2,e3) ->
      (
        match eval env e1 with
            V_Bool true   -> eval env e2
          | V_Bool false -> eval env  e3
          | _ -> raise TypeError                                 )

  | E_Get t -> store_get t !store 
  | E_Set (t,e) -> 
      let v = eval env e in 
        begin store := store_set t v !store ; v end  
\end{verbatim}

In OCaml, the expression \texttt{!r} is a read access to the
value of the reference \texttt{r}, the expression \texttt{r := e} 
sets the value of \texttt{e} into the reference \texttt{r}
and the expression \texttt{begin e1 ; e2 end} is a sequential evaluation of \texttt{e1} and \texttt{e2} which returns the value
computed for \texttt{e2}.

That code is finally the  same as in section~\ref{sec-classical}
with two additional cases for \emph{get} and \emph{set}
which make use of the store as a mutable global variable.

Whether this version is better than the previous one or not is not the subject of this paper. However we can compare them.
The first one is more easy to formalize since it does not
use imperative features, but the store is present in all the
parts of the code. 
The second one is more difficult to formalize but it is
simpler to read, as long as the imperative part is well
understood.
In that one, imperative features and purely functional
features are completely independent which shows a good
separation of concerns and makes easier the extension of the
language with new features.

This point is particularly important for us since our goal is to
build an easily extendable interpreter. 

For these reasons, we will use such a mutable global
variable to represent the store in the implementation of our
scheme of interpretation and we will keep the imperative
implementation of the classical evaluator as a reference to
compare the two approaches.

\subsection{Imperative features in our scheme}
\label{sec-imperative-scheme}

In section~\ref{sec-scheme}, when building the variable
elimination algorithm (step $\mathcal C$), we have rejected rules which were
correct in the call-by-name setting but false in the
call-by-value setting. 
By doing so, we have ensured that the order of the
applications was not disturbed by the transformation $\mathcal C$.

This means that as long as imperative features are triggered
by the evaluation of an application, the transformation
$\mathcal C$ preserves the order of the imperative features.

This means also that it is sufficient to represent features
by function applications to integrate them nicely in our
interpretation scheme.

For this reason, in our scheme we transform all imperative
features into applications as we have done for non-strict
features.
The imperative features are then embedded in external
functions with side effects. 

Here is the code doing it as an extension of
the \emph{purify} function (step $\mathcal{P}$).

\begin{verbatim}
let rec purify = function 
 ...
   | E_Get m -> P_App (make_get m , dummy_val)
   | E_Set (m,e) -> P_App (make_set m, purify e) 
\end{verbatim}

\begin{verbatim}
let make_get t = 
   P_Const (V_Fun (fun _ -> store_get t !store)) 

let make_set t = 
   P_Const (V_Fun (fun v -> begin  
                             store := store_set t v !store ; 
                             v
                            end))
\end{verbatim}

This is sufficient to deal with imperative
features since having the store as a global variable allows
to keep the rest of the code unchanged as seen in
section~\ref{sec-classical-imperative}.
The other steps $\mathcal C$ and $\mathcal E$ are left
unchanged.

This example shows that adding new features in our language
may be very simple since it can be sufficient to extend only
one step. Furthermore, extending that step is rather simple
since it is done by adding one case to the transformation.

\section{Experimental evaluation}
\label{sec-performances}

In this section, we observe the performances of our
interpretation scheme. 
We first compare it to the classical implementation
proposed, then to widely used interpreters for other
(dynamically typed) languages.

\subsection{Comparison to the classical scheme}
We compare execution times of our interpreter to execution
times of the classical interpreter by running both of them
on several source programs.
The set of source programs we have selected is composed of
the Fibonacci and Ackermann functions, an insertion sort and
an implementation of the 8 queens problem.

For Ackermann, Fibonacci and sort, we provide two different
source codes. In the first one, the \emph{omega} function
$\lambda x. (x\,x)$ is used to encode recursive calls (see
Figure~\ref{fibo-sources}). In the other one, imperative
references are used to do this.
In order to implement the sort algorithm and the 8 queens
problem, we have added lists in the interpreted language :
lists of values are added to the \texttt{value} type and the
list operators (cons, head, and tail) are external
functions.
The corresponding input programs and interpreters are
provided with this report.

We give the results for two different variable elimination
algorithms noted $\mathcal C_1$ and $\mathcal C_2$ (or
$\mathsf C1$ and $\mathsf C2$ in diagrams). The first one,
$\mathcal C_1$, has the following features: \begin{itemize}

\item use of the combinators $\combis$, $\combik$, $\combii$, $\combib$, $\combic$, $\combin$ and $\combisif$ (sections~\ref{sec-BCN} and~\ref{sec-sif}),

\item use of the automatic pre-evaluation (section~\ref{sec-pre-eval}). 

\end{itemize}

This algorithm has been chosen because it has a good ratio
performances/coding effort (using only $\combis$, $\combik$ and $\combii$ gives too poor performances).

The second algorithm, $\mathcal C_2$,  has the following features:
\begin{itemize}

\item $\combis$, $\combik$, $\combii$, $\combib$, $\combic$, $\combin$,

\item $\combis_2$, $\combis^2$ and $\combis^2_2$ (section~\ref{sec-n-ary}), 

\item all combinators extending $S_2$ to the selective distribution of the abstractions ($\combib$, $\combic$, $\combin$ style),

\item all combinators extending $S^2$ to the selective distribution of the abstractions ($\combib$, $\combic$, $\combin$ style),

\item $\combisif$ and ${\combisif}^2$ (in the style of $\combis^2$),

\item use of the automatic pre-evaluation (section~\ref{sec-pre-eval}). 

\end{itemize}

This algorithm is the best we have found with the presented optimizations without using all the combinations of $S_2^2$ with the selective distribution in the $\combib$, $\combic$, $\combin$ style (which corresponds to Diller's combinators~\cite{diller02}).

The following diagram shows the execution time of our two
implementations based on $\mathcal C_1$ and $\mathcal C_2$ normalized against
 the classical interpreter (section~\ref{sec-classical}) on several input programs with
short execution times (less than 1 second for the classical
interpreter). 
The machine used is a Pentium D 2.8 GHz with 1 GB of memory, running Linux.

\begin{center}
\includegraphics[width=12.5cm]{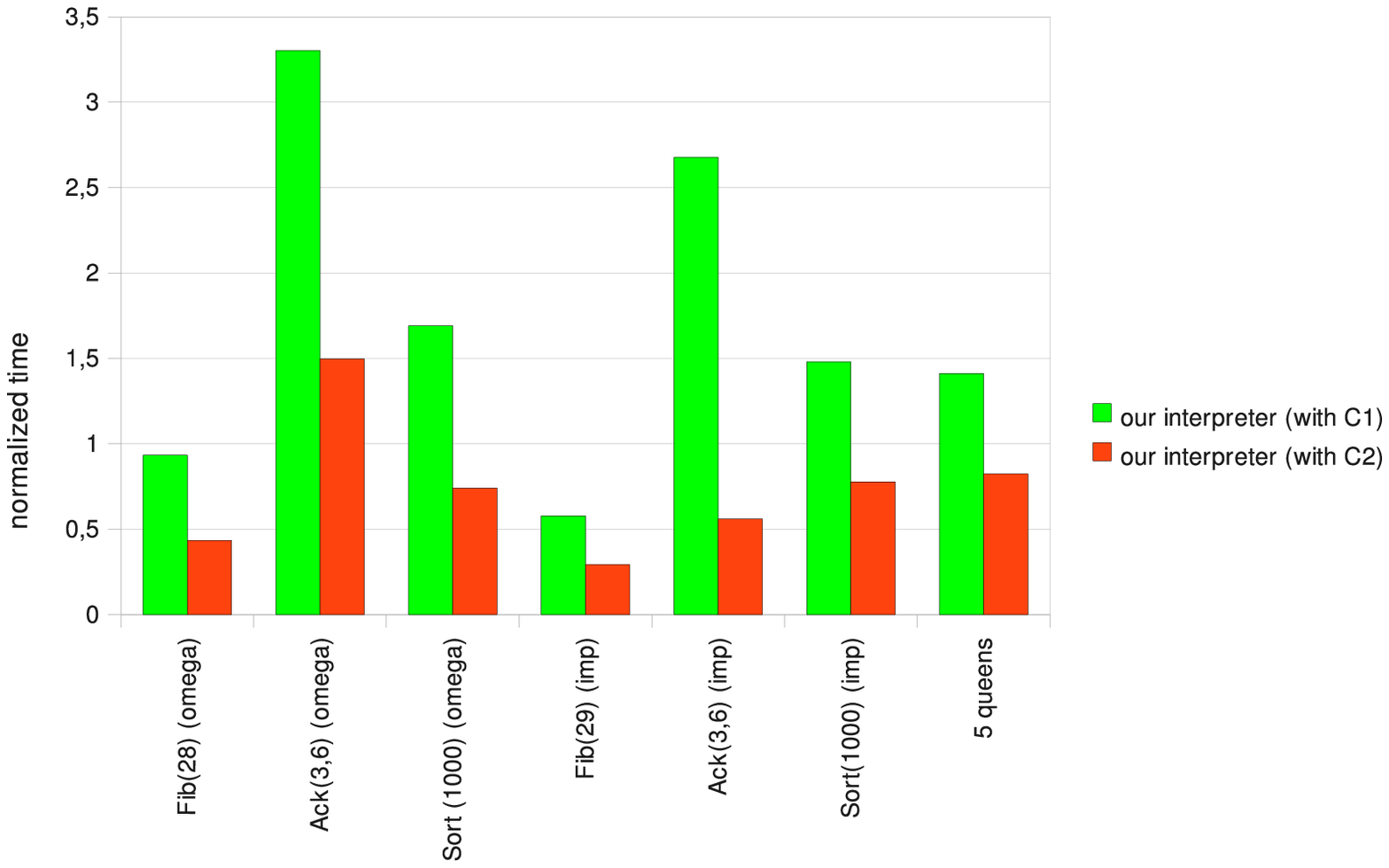}
\end{center}

We see that $\mathcal C_1$ has erratic
performances. Performances of $\mathcal C_2$ are also
irregular, but better than the classical interpreter on
most of these inputs.

\begin{figure}[!htp]

\includegraphics[width=6cm]{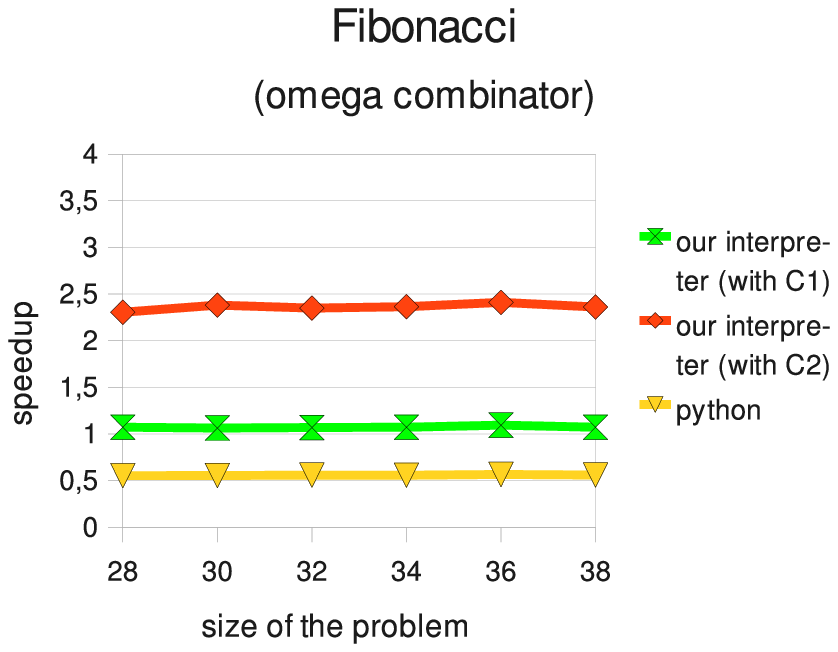}
\hfill
\includegraphics[width=6cm]{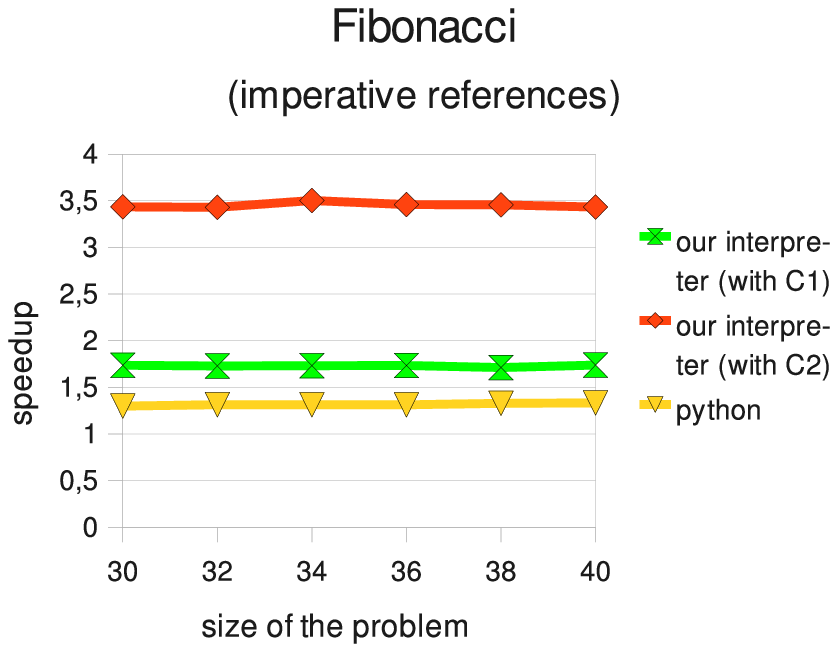}

\includegraphics[width=6cm]{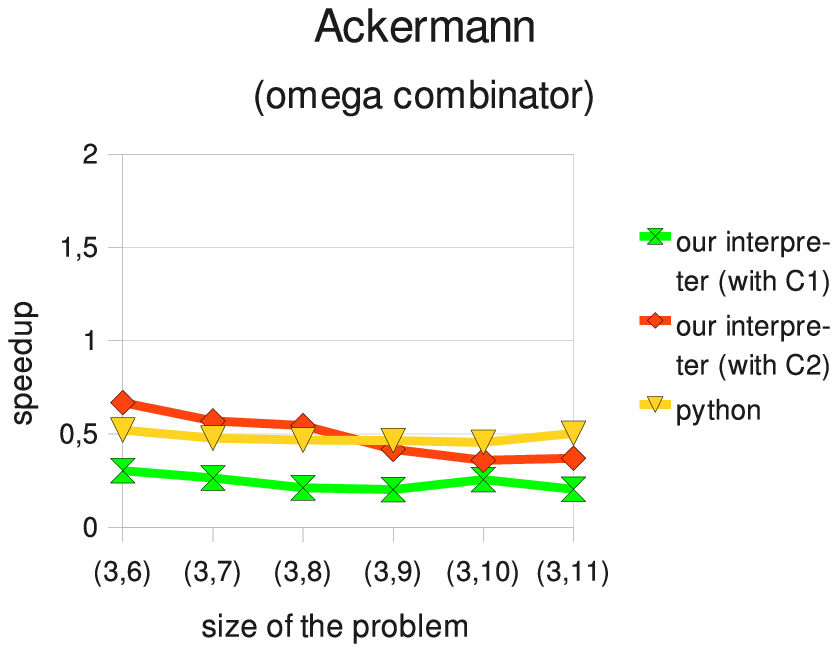}
\hfill
\includegraphics[width=6cm]{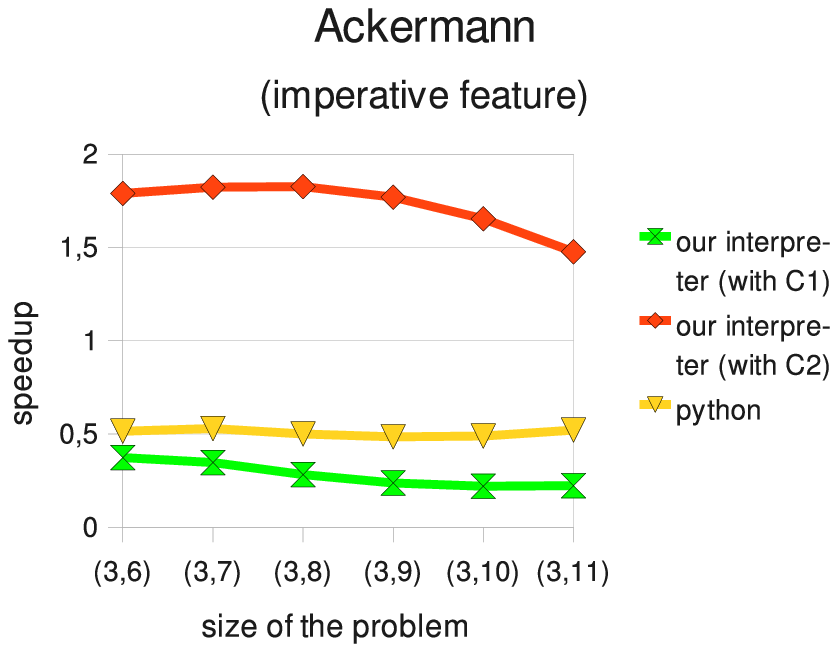}

\includegraphics[width=6cm]{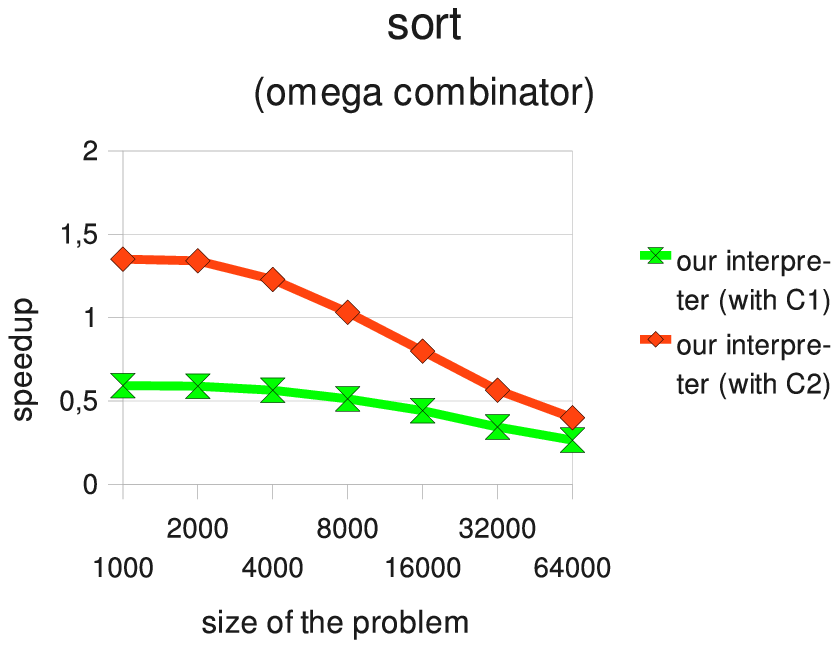}
\hfill
\includegraphics[width=6cm]{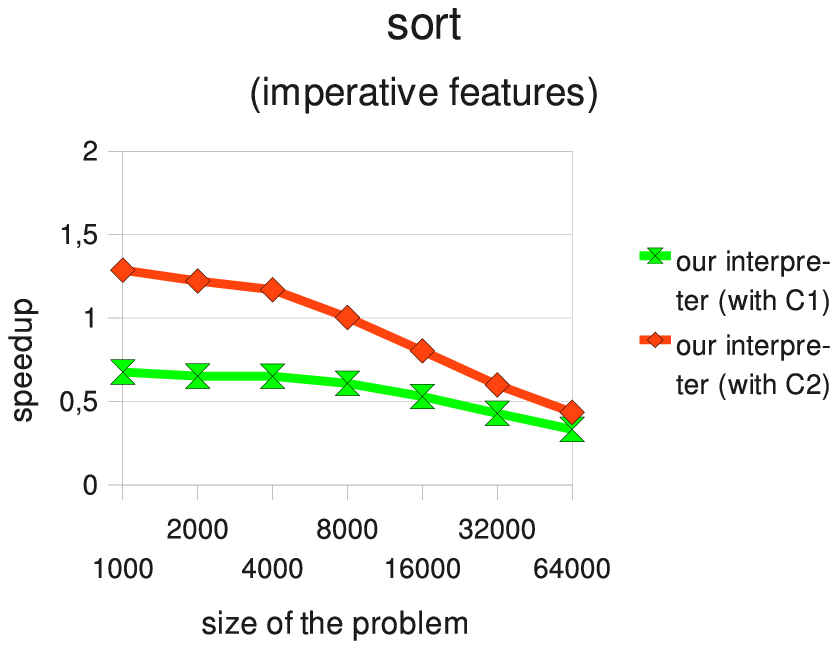}

\hfill\includegraphics[width=6cm]{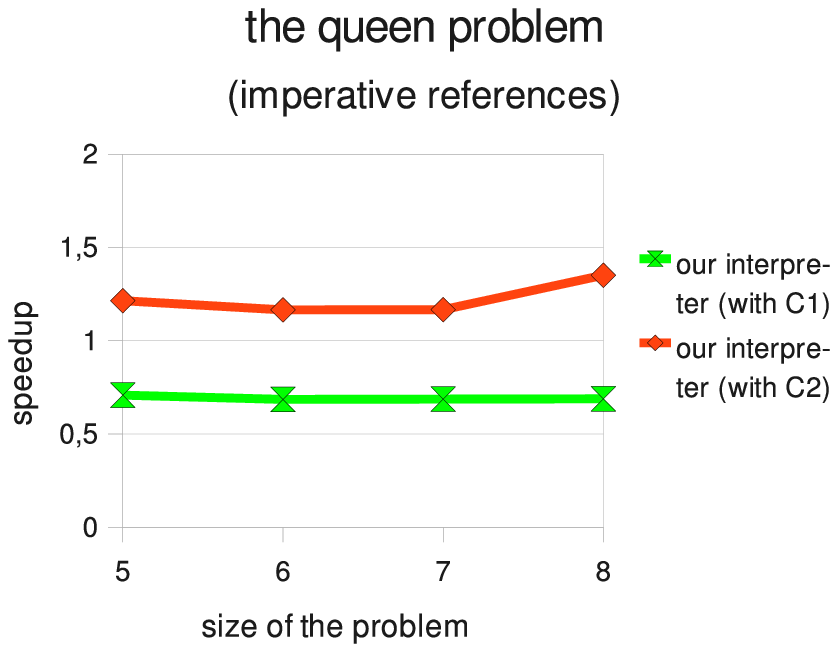}

\caption{Speedups for two variable elimination algorithms}
\label{fig-input-size}

\end{figure}

\begin{table}[!htp]
\begin{center}
\begin{tabular}{|ll||l||l|l||l|l|}
\hline
                     &         & Cl.  & $\mathcal C_1$ & $\mathcal C_2$ & Python & Erlang \\
\hline
Fibonacci (omega)    & 28      & 0,77  & 0,72           & 0.34           & 1.39   & 341   \\
Fibonacci (omega)    & 38      & 94    & 87,67          & 39,82          & 167,94 &       \\
Fibonacci (imp)      & 29      & 0.97  & 0.56           & 0.28           & 0,76   & \\
Fibonacci (imp)      & 40      & 195,0 & 112,2          & 56,8           & 146,3  & \\
Ackermann (omega)    & (3,6)   & 0.34  & 0.86           & 0.39           & 0.50   & 146 \\ 
Ackermann (omega)    & (3,11)  & 357.9 & 1744           & 967            & 711    & \\ 
Ackermann (imp)      & (3,6)   & 0.17  & 0.46           & 0.10           & 0.33   & \\
Ackermann (imp)      & (3,11)  & 254   & 1139           & 172            & 488.0  &\\
sort (omega)         & 1000    & 0,79  & 1.34           & 0.59           & /      & \\
sort (omega)         & 64\,000 & 5603  &                & 14046          & /      & \\
sort (imp)           & 1000    & 0.61  & 0.90           & 0.47           & /      & \\
sort (imp)           & 64\,000 & 4516  & 13570          & 10381          & /      & \\
queens problem (imp) & 8       & 761   & 1105           & 563            & /      & \\ 
\hline
\end{tabular}
\end{center}
\caption{Performances (in seconds, on Pentium D)}
\label{table-times}
\end{table}

Now, let us have a look at the performances when the size
of the problem increases. 
The diagrams of figure~\ref{fig-input-size} show
 speedups for increasing problem sizes. Reference times are given in Table~\ref{table-times}.

\newcommand{\compbcn}{\mathcal{C}_1}
\newcommand{\compopt}{\mathcal{C}_2}

For Fibonacci and the 8 queens, the performances are stable. 
For the sort, they are decreasing when the size of the problem increases (the number of elements to sort).
Our interpreter based on~$\compopt$ becomes slower than the classical interpreter for 8000 elements, when the execution time is 57 seconds (omega) or 45 seconds (imperative references).
On Ackermann (omega) we observe bad performances which are getting worth when the size of the problem increases.
Note that we can find combinations of optimizations giving
better performances for this input program, but the ones we
have considered have made decrease the performances for
less specific input programs such the sort and the 8 queens.
Ackermann (imperative references) shows also decreasing speedups, but it stays above 1 in our experiments.

\begin{figure}[!htp]
\includegraphics[width=6cm]{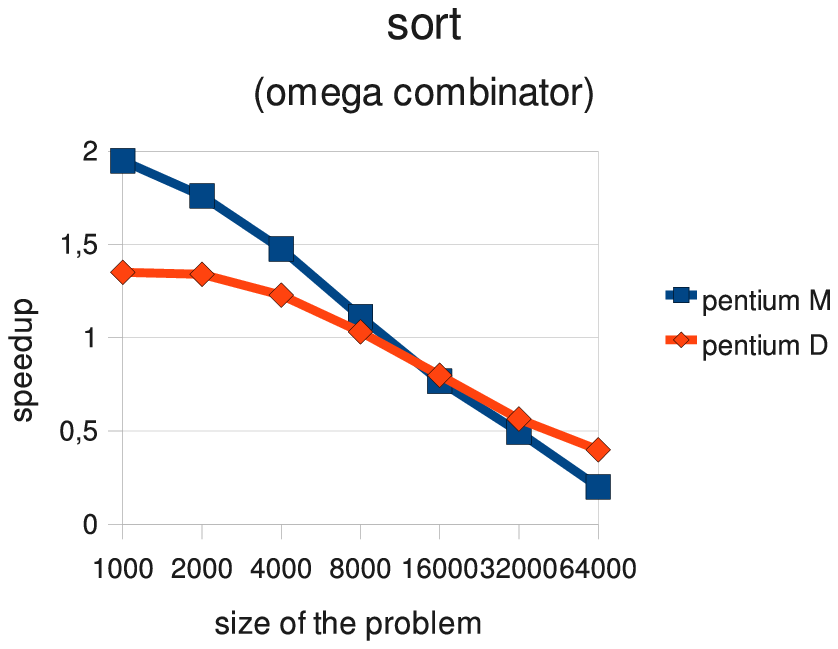}
\hfill
\includegraphics[width=6cm]{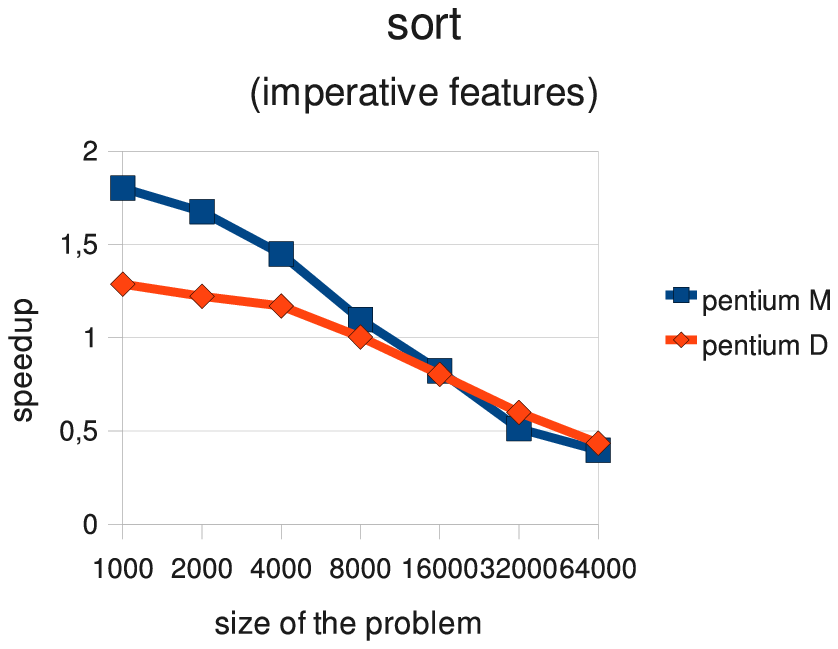}


\includegraphics[width=6cm]{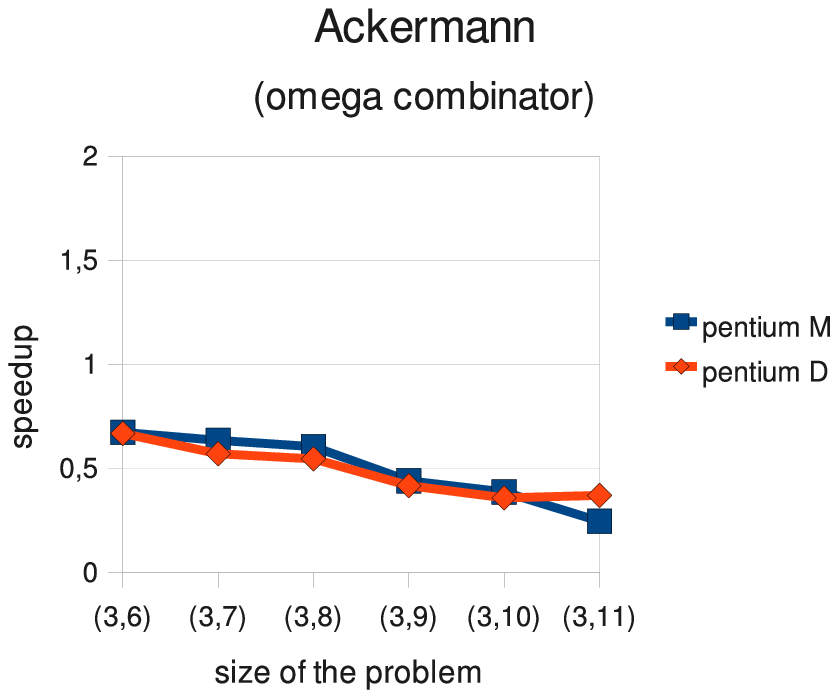}
\hfill
\includegraphics[width=6cm]{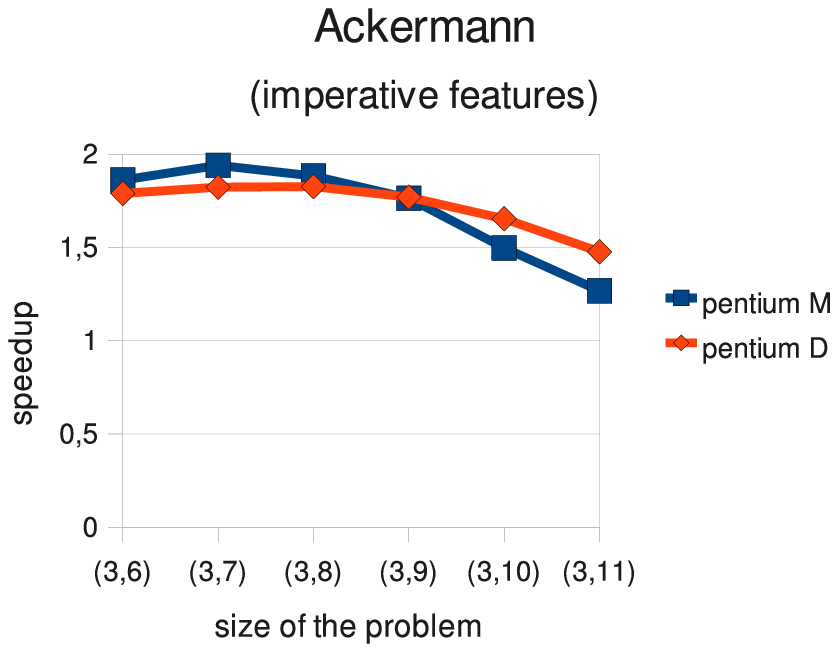}


\hfill\includegraphics[width=6cm]{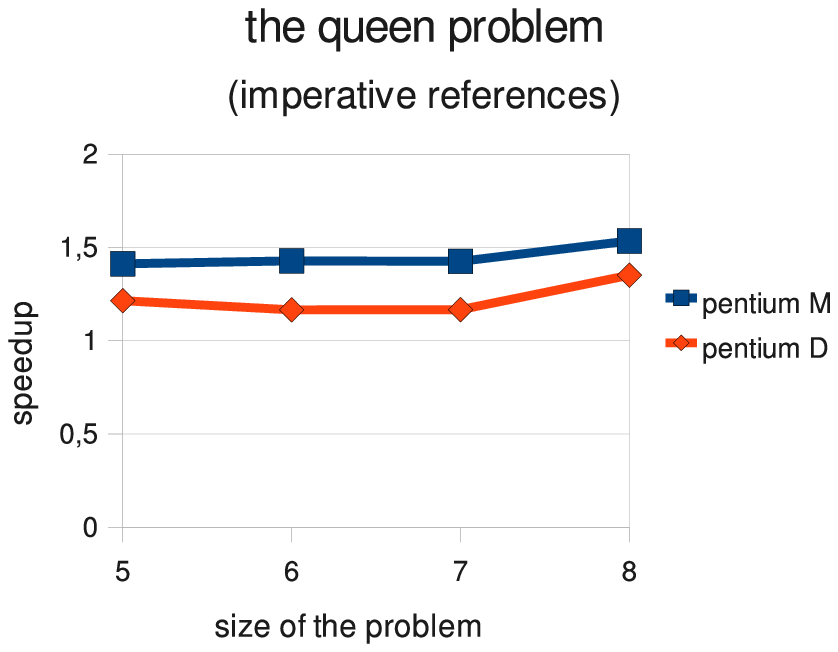}

\caption{Speedups for our best algorithm on several programs, on two different hardwares}
\label{fig-comparison-architectures}
\end{figure}

Figure~\ref{fig-comparison-architectures} compares the
speedups of $\mathcal{C}_2$ on two different architectures :
the Pentium D already used for previous experiments, and a 
Pentium M (laptop) 1.1 GHz with 512 MB of memory, also
running Linux.

For the sort, the speedup is better on the laptop for small problems and it tends to be equal or worth than the desktop machine as the speedup becomes a slowdown.
For the queen problem, the difference between speedups is
stable (between 14\% and 23\% of difference), but for Ackermann, it is not (50\% for Ackermann(3,11) (omega)).

During all our experiments, no memory leak has been observed.
Furthermore, the technique discussed in this paper has been
used in the MGS interpreter, which has been used for
intensive biological simulations~\cite{godin2006} running
for days, and no memory leak has been observed either.

\subsection{Comparison to widely used interpreters}

Now, let us compare our interpreter with other interpreters which are widely used.
Of course, comparing the performances of our interpreter to
the ones of statically typed languages such as OCaml or Java
tends to show that our interpreter has poor performances, but it
is fairer to compare  to dynamically typed languages.

Performances of the Python interpreter (2.6.4) are given in
Figure~\ref{fig-input-size} and table~\ref{table-times} for
our programs that do not use lists (lists in Python are
modified in place so we could not use the same input
programs).

The source codes for Python programs used in these tests are
as close as possible to the code used as input of our
interpreters (see for instance the code for Fibonacci in Figure~\ref{fibo-sources}). 
In particular, curried $\lambda$-expressions are used. The
Python interpreter shows much better performances with
uncurried functions, but this would be also the case for
the classical interpreter and in our scheme of
interpretation.

Our comparison shows that the performances of our
interpreter can compete with Python's ones (Python 2.6.4).
 We conclude that the performances of our scheme, even if it
 does not improve the performances of the classical, naive
 interpreter, are acceptable.


 Table~\ref{table-times} gives also some results for the
 Erlang interpreters.
The very poor performances observed for Erlang can be
explained by the fact that our test programs use mainly
anonymous, uncurried functions(Fig.~\ref{fibo-sources}), which are available in Erlang
but are not designed to be the main means of defining
functions.

\begin{figure}[ht]

\hrule

\medskip

Fibonacci in our language:\\

\begin{boxedminipage}{\textwidth}
\begin{verbatim}
(((\f.(f f))
  (\f. \n. if (n <= 1) 
           then 1 
           else (((f f) (n - 2)) + ((f f) (n - 1))) fi)
 ) 28)
\end{verbatim}
\end{boxedminipage}\\

\medskip

Fibonacci in Python:\\

\begin{boxedminipage}{\textwidth}
\begin{verbatim}
(((lambda x : x(x))
 (lambda f : lambda n : (1 if (n <= 1)
                         else ( (f)(f)) (n-2) + (f(f)) (n-1) )))                  
 (28))
\end{verbatim}
\end{boxedminipage}\\

\medskip

Fibonacci in Erlang:\\

\begin{boxedminipage}{\textwidth}
\begin{verbatim}
((fun (F) -> F (F) end) 
 (fun (F) ->
      fun (N) -> 
              if (N =< 1) -> 1 ;
                 true -> (((F (F)) (N - 2)) + ((F (F)) (N - 1))) 
              end end end))
 (28).
\end{verbatim}
\end{boxedminipage}

\caption{The Fibonacci function using the omega function in our language, Python and Erlang.}
\label{fibo-sources}

\end{figure}

\clearpage

\section{Correctness}
\label{sec-correctness}

\newtheorem{theorem}{Theorem}
\newtheorem{definition}{Definition}

We have seen in section~\ref{sec-ski} that some classical
variable elimination algorithms are not correct in the
call-by value setting.
However, the algorithm~$(fa_vb)$ is known to be correct in
the $\lambda_v$-calculus~\cite{cbv-combinatory-logic}.
We provide a mechanically checked proof that the
algorithm~$(fa_vb)$ is also correct when we add some
imperative features into that language.

The proof is checked with the Coq proof assistant~\cite{coq} and is
provided as Coq files with this report. In this
section we explain what is proven rather than the proof
itself.

In our language,
combinators are seen as particular $\lambda$-abstractions to simplify the proof.
The property we show is that a term $e$ and the term
$\mathcal C(e)$, which is the transformation of $e$ by the
$(fa_vb)$ algorithm, share the same semantics.
This idea is formalized by an equivalence relation.

Some definitions we give here are extended from those
of~\cite{plotkin} or~\cite{cbv-combinatory-logic}. 

\newcommand{\mget}{\Gamma}
\newcommand{\mset}{\Sigma}
\newcommand{\reduces}{\longrightarrow}
\newcommand{\subst}[3]{[#2/#1]#3}
\newcommand{\dans}{~/~}

\paragraph{Terms.}

\newcommand{\langage}{\Lambda_{vr}}

We consider the language $\langage$ defined below. 
We do not consider constants, external functions and
conditionals, but we take imperative
references into accounts. $\mathit{Tags}$ is a set of tags,
$\mathit{Vars}$ is a set of variables, $\mathit{Vals}$ is
the set of values of the language.

\begin{definition}[syntax of $\langage$]

\begin{align*}
  M & \in \langage\\
  a_f & \in \mbox{\it{Fun-Consts}}\\
  x & \in \mathit{Vars}\\
  V & \in \mathit{Vals}
\end{align*}
\begin{align*}
 V &::= ~ x ~|~ \lambda x.M ~|~ a_f\\
 M &::= V ~|~ (M ~  M) 
\end{align*}

\end{definition}

Only applications are not values.
We follow Plotkin's $\lambda_v$-calculus~\cite{plotkin} and
 consider variables as values.  This is convenient since the
 semantics allows variables to be substituted only for
 values. Moreover, in practice, programs with free variables
 will be rejected before being evaluated. Taking variables
 as values simplifies the proof.

The construct $a_f$ corresponds to external
functions. 
Imperative features are seen as external
functions.
 For instance, the reference operators seen in
section~\ref{sec-imperative} are seen as particular external
functions (we can consider there is an operator for each
reference).

\paragraph{Stores.}

We consider an abstract set of stores $S$.
That set comes with a set $F_S$ of total functions of type
$ \mathit{Vals} \times S \rightarrow   \mathit{Vals} \times S$.
For instance, \codeline{store\_get t}
and \codeline{store\_set t} (for a given \codeline{t}) of
section~\ref{sec-imperative} can be viewed as such functions
(by adding dummy parameters and results to their
original type).

A total function $\delta : \mbox{\it{Fun-Consts}} \rightarrow F_S$ gives the semantics of the external functions of the language.
This function is also left abstract.

We do not need to enforce more properties on stores for the proof. 
In the following  $\sigma$ stands for a store (an element of $S$).

\paragraph{Reduction.}

The relation $\reduces$ is defined inductively by the rules
below. This relation gives the call-by-value operational
semantics of the language.

\begin{definition}[reduction]

$$\frac{}{((\lambda x.M) ~ V) \dans \sigma \reduces \subst x V M \dans \sigma}(\beta_{vr}) $$


$$\frac{M_1 \dans \sigma_1 \reduces M_2 \dans \sigma_2}
{(M_1 ~ M) \dans \sigma_1 \reduces (M_2 ~ M) \dans \sigma_2}(\mathit{Left})$$


$$\frac{M_1 \dans \sigma_1 \reduces M_2 \dans \sigma_2}
{(V~M_1) \dans \sigma_1 \reduces (V~M_2) \dans \sigma_2}
(\mathit{Right})$$


$$\frac{(V_2,\sigma_2) = \delta(a_f)(V_1,\sigma_1)}
{(a_f~V_1) \dans \sigma_1 \reduces V_2 \dans \sigma_2}(\delta)$$

\end{definition}

The $\beta_{vr}$ rule corresponds to the usual $\beta$ reduction rule in
call-by-value $\lambda$-calculus : 
reduction can be triggered only when the argument is a
value. We use the  classical proper substitution with automatic
variable renaming (we do not adopt Barendregt convention
requiring renaming to avoid collisions). 
This difference between the proof setting and our
interpreter is not important since \emph{in practice} we are
interested in programs with no free variables 
and then the interpreter does not need to rename variables (values used
in $\beta$-reductions never contain free variables).

The \emph{Left} and \emph{Right} rules state that we 
reduce the left hand side of an application before
reducing the right hand side.

The $(\delta)$ rule associate a semantics functions to
external function symbols. In order to take imperative
features into accounts, semantics function of $F_S$ take the
``current'' store as a parameter and the returned store is
set as the ``current'' store. 
This corresponds to the monad style, whereas in
section~\ref{sec-imperative-scheme} the store was a global
mutable variable.
As discussed in section~\ref{sec-classical-imperative}, this
should make no difference in the call-by-value setting.

\paragraph{Equivalence.}

\newcommand{\rel}{\simeq}

We define a relation of equivalence on~$\langage$
noted~$\rel$.
The rules are based on the ones for the $\lambda_v$-calculus
in~\cite{cbv-combinatory-logic}.

\begin{definition}[term equivalence]

$$\frac{}{M \rel M}(\mathit{reflexivity}) \quad \quad \quad \quad \frac{M \rel N}{N \rel M}(\mathit{symmetry})$$

$$\frac{M_1 \rel M_2 \quad M_2 \rel M_3 }{M_1 \rel M_3}(\mathit{transitivity})$$

$$\frac{M_1 \rel M_2}
{(M_1 ~ N) \rel (M_2 N)}(\mbox{\it{Compatibility-L}}) 
\quad \quad
\frac{M_1 \rel M_2}
{(N ~ M_1) \rel (N  ~ M_2)}(\mbox{\it{Compatibility-R}})$$


$$\frac{\forall \sigma, M_1 \dans \sigma \reduces M_2 \dans \sigma}{M_1 \rel M_2}(\beta_{vr}\mathit{-eq})
\quad \quad \quad \quad
\frac{M_1 \rel M_2}{\lambda x . M_1 \rel \lambda x . M_2}(\xi)$$

$$\frac{(\lambda x_1.M_1) ~ y \rel (\lambda x_2.M_2) ~ y \quad
 y \mbox{ not free in } \lambda x_1.M_1  \quad
 y \mbox{ not free in } \lambda x_2.M_2}
{\lambda x_1.M_1 \rel \lambda x_2.M_2}(\zeta')$$

\end{definition}

\paragraph{Combinators.}
We consider the combinators as particular
terms of the language : $\combii = \lambda x.x$, $\combik = \lambda x. \lambda y.x$ and $\combis  = \lambda x.\lambda y. \lambda z . ( (x\,z)\,(y\,z))$,  where $x$, $y$ and $z$ are distinct elements of
$\mathit{Vars}$ (these are not meta-variables).

\newcommand{\distrib}{\mathcal D} 
\newcommand{\combi}{\mathcal C}

\paragraph{Variable elimination.}
We now define the $(fa_vb)$  algorithm, noted $\mathcal C$.

\begin{definition}[variable elimination algorithm] The functions $\combi$ and $\distrib$ are defined inductively by the following rules ($x$ and $y$ range over $\mathit{Vars}$) :

\begin{align*}
\distrib (x, x)& = \combii \\
%
%
\distrib (x,  (M_1 ~ M_2)) & = ((\combis~\distrib (x, M_1)) ~ \distrib(x, M_2))) \\
%
%
%
%
%
%
\distrib (x,y) & = (\combik~y) \mbox{ where } x\not = y \\
%
%
%
\distrib(x,c) & = (\combik~c) \mbox{ where } c \in \{\combii,\combik,\combis,a_f\}
\end{align*}

\begin{align*}
\combi(x) & = x\\
%
\combi(\,(M_1 ~ M_2)\,) & = (\combi(M_1) ~ \combi(M_2))\\
%
\combi(\lambda x.M) & = \distrib(x,\combi (M))\\
\combi(a_f) &= a_f
\end{align*}

\end{definition}

\paragraph{Correctness.} The correctness theorem states that a term and its transformed form are equivalent.

\begin{theorem}[Correctness of $(fa_vb)$]
$\forall M \in \Lambda_{vr}, M \rel \combi(M)$
\end{theorem}

This result means that, for a given store, if a term $M$
evaluates into a value $V$ then $\mathcal C (M)$ evaluates
into a value which is equivalent to $V$.
Furthermore, if the evaluation of $M$ is stuck then the same is true for
the evaluation of $\mathcal C(M)$ (but not necessarily in
an equivalent store) and if it does not terminate, so for
$\mathcal C(M)$.

The proof of this theorem is checked by the Coq
system~\cite{coq}, the corresponding Coq files are provided
with this report.

\section{Choice of the host language}

\label{sec-host-language}

We have illustrated our scheme with OCaml as host
language. 
However, our scheme can be used with other host languages as long
as they can deal with functions and they  have a call by
value reduction strategy.

In C language, functions can be handled with pointers. Since closures are necessary for managing partial applications (for $\combik$ and $\combis$ for instance), it is necessary to encode closures, for instance with records, and a form of garbage collecting.
In C\#, lambda expressions are available directly in the
language and in C++ as a library.
 In Java, function are encapsulated into objects (or
 classes). 

In this section, we give guidelines to use of our scheme in
 Java. 
In particular, we give the higher order data structure
needed for the evaluation step ($\mathcal E$), after what
the three steps $\mathcal P$, $\mathcal C$ and $\mathcal E$
become trivial to implement following 
section~\ref{sec-scheme}.

\subsection{The \codeline{Value}  data structure}
We give here the data structure to represent the
type \codeline{value} of section~\ref{sec-scheme}.
We adapt the composite pattern which is usually used to
represent abstract syntax trees in object languages.
At the root of the hierarchy, we find an abstract
class: \codeline{Value}.

\begin{verbatim}
public abstract class Value {}
\end{verbatim}

\begin{center}
\includegraphics[width=7cm]{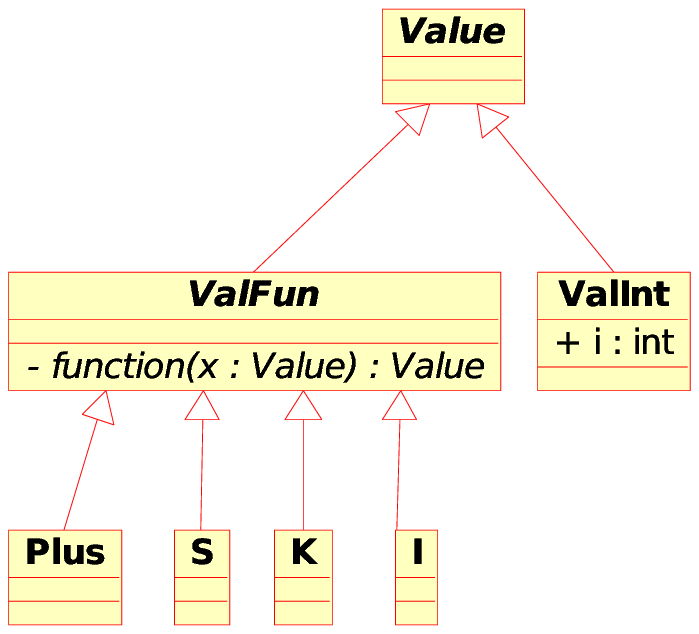}
\end{center}

The most basic sort of values are integers (we skip booleans here):

\begin{verbatim}
public final class ValInt extends Value {
    public int i ;
    public ValInt (int i){
        this.i = i ;
        }
    }
\end{verbatim}

We see functions as objects with a method
called \codeline{function}, taking a \emph{value} as argument and
returning a \emph{value}.
Since in Java all the objects of a class have the same
methods (functions are not first-class values), we 
define a class for each different function. For this reason, the
class for functions is abstract.

\begin{verbatim}
public abstract class ValFun extends Value {
    abstract Value function (Value x) ;
    }
\end{verbatim}

\subsection{Pre-defined functions and combinators}

We now  define some combinators and an external function as subclasses of
 \codeline{ValFun}. 
The combinator $\combii$ is direct:

\begin{verbatim}
public class I extends ValFun {
    Value function (Value x) { return (x) ; }
    }
\end{verbatim}

The combinator $\combik$ has two arguments. 
So we have to use closures since the two arguments are not
passed at the same time.
Closures are usually encoded in Java with internal classes.
Here, the method returns an object of a new subclass
of \codeline{ValFun} at each call:

\begin{verbatim}
public class K extends ValFun{
    Value function (final Value c) {
        return new ValFun(){
            Value function (Value dummy) { return (c) ; }
            };
        }
    }
\end{verbatim}

The same is done for the combinator $\combis$ with three arguments:

\begin{verbatim}
public class S extends ValFun {
  Value function (final Value f) {
    return new ValFun(){
      Value function (final Value g) {
        return new ValFun(){
          Value function (Value x) {
             ValFun fx = (ValFun) ((ValFun) f).function(x) ;
             Value  gx =          ((ValFun) g).function(x) ;
             return (fx.function(gx));
             }
          };
        }   
      };
    }
  }
\end{verbatim}

Let us now consider the integer addition as an example of external function. We suppose \codeline{ExternalFuns.plus} is the usual operator on base integers.
It has two arguments so the method uses closures:

\begin{verbatim}
public class Plus extends ValFun{
  Value function (final Value x) {
    return new ValFun(){
      Value function (final Value y) {
         int vx = ( (ValInt) x ).getInt();
         int vy = ( (ValInt) y ).getInt();
         return( new ValInt (ExternalFuns.plus(vx, vy) ));
         }
      };
    }
  }
\end{verbatim}

\section{Conclusion}
\label{sec-conclusion}

\subsection{Benefits and drawbacks of our approach.}

\paragraph{Initial Constraints verified.}
The first advantages have been put as conditions : \begin{itemize}

\item Mainstream languages as Java or C can be used.

\item The scheme supports higher-order functions, non-strict
  features, imperative features and external functions.

\item The performances of the interpreter are acceptable
  compared to other widely used interpreters (for dynamically typed languages) and can be improved with standard optimization such as uncurrying of functions.

\end{itemize}

\paragraph{Unifying external and user-defined functions.} 
We have seen that we transform user-defined functions into
host language functions. External functions are also
represented by host language functions. A consequence of
that unification is that the handling of higher order is
very simple, even when using function in libraries defined
independently of the interpreter. 
For instance, in order to evaluate our interpreter on
programs such as sorting and the queen problem, we have
included lists in the language. Higher order functions such
as \codeline{List.filter} from the OCaml standard library
have been included in the language with only structural wrappers
to deal with dynamical typing. In the classical approach,
the wrapper has to check the kind of function received, and
when a user-defined function is received, then a call to
\codeline{eval} is needed. Unlike the wrapper in our
approach, the wrapper is not only structural anymore and it
depends on the evaluation function.

\paragraph{Simplicity of the scheme and of the evaluator.} 
We have seen that the evaluation step $\mathcal E$ in the scheme is extremely simple. The step $\mathcal P$ which transforms non-strict and imperative features into functions is also simple. Depending on the number of optimizations needed, the step $\mathcal C$ which transforms user-defined functions into combinators can become complex. 
However, the algorithm keeps cleanly specified by a set of
simple rewriting rules. That algorithm can be implemented
modularly by structural pattern matching (a rewriting rule
in the algorithm becomes a pattern-matching case in the
implementation) or other tree traversal technique such as in
the Composite design pattern. The most complex algorithm we
have used ($\mathcal C_2$ in section~\ref{sec-performances}) is about 100 lines long (plus 200 lines for the definition of the 34 combinators).

Moreover, each step is independent of the other, and many new features can be added into the interpreted language without modifying $\mathcal C$ or $\mathcal E$, just by adding a case to~$\mathcal P$.

\paragraph{Traversal is difficult.}
The recurrent drawback of higher-order abstract syntax is
the difficulty to traverse the syntax trees.
Indeed, host language functions cannot be inspected as
ordinary trees can. This means that once the user-defined
functions have been transformed into host language
functions, further analysis or transformations are very
limited.
Solutions to this problem are not simple,
see~\cite{boxes-go-bananas} for instance.
However, many analysis and transformations can still be done
before transformation into combinators.
Moreover, in our higher order abstract syntax trees, if we
except the pre-evaluation optimization, only a few number of
pre-defined host language functions are encountered, and
they can be identified by using physical equality for
instance.

\subsection{Related work}

Although the idea is not new (it is suggested
in~\cite{boxes-go-bananas}), higher-order abstract syntax
has found few applications for building interpreters.
The two first experiments the authors are aware of is the
implementation of the MGS language which this paper is based
on~\cite{these-Julien}, and the work of
Barzilay~\cite{Barzilay2006}. 
The latter makes the
assumption that the host language has some multi-level
programming facilities, such as generating new host-language
variables at execution time, which we do not consider since we want to be able to use languages such as Java or C.
This difference is important only for the step $\mathcal C$
which transforms user-defined functions into host language
functions. With multi-level programming facilities, it is
not necessary to decompose functions into combinators.
Note also that Barzilay deals with call-by-name and
call-by-need whereas we deal with call-by-value strategy.

The fact that the generation of
host language functions is not trivial and that the
generated functions cannot be inspected might explain that
higher-order abstract syntax had not been used before to implement
interpreters.

\bibliographystyle{alpha}

\begin{btSect}{websites}
\section*{Web Sites}
\btPrintCited
\end{btSect}

\begin{btSect}{references}
\section*{References}
\btPrintCited
\end{btSect}

\end{document}